\documentstyle[onecolumn]{mn}
\newcommand{\reference}{\bibitem}
\def\beq{\begin{equation}}
\def\eeq{\end{equation}}
\def\bey{\begin{eqnarray}}
\def\eey{\end{eqnarray}}
\def\beqarray{\begin{eqnarray}}
\def\eeqarray{\end{eqnarray}}

\def\cm{\,{\rm {cm}}}

\def\kpc{\,{\rm {kpc}}}
\def\pc{\,{\rm {pc}}}
\def\kpch{\,{h^{-1}{\rm kpc}}}

\def\kms{\,{\rm {km\, s^{-1}}}}
\def\msun{\,\rm M_\odot}

\def\v200{V_{200}}

\def\Rd{R_{\rm d}}

\def\Md{M_{\rm d}}

\def\md{m_{\rm d}}

\def\mtold{\Upsilon_{\rm d}}
\def\lsun{\,\rm L_\odot}
\def\Lsun{\,\rm L_\odot}
\def\lsunI{\,{\rm L_\odot}_I}

\def\jd{j_{\rm d}}

\def\rmd{{\rm d}}
\def\rmh{{\rm h}}
\def\rms{{\rm s}}
\def\calM{{\cal M}}
\def\calI{{\cal I}}
\def\calK{{\cal K}}

\input epsf
\title[]        
{Tully--Fisher Relation and its Implications for 
Halo Density Profile and Self-interacting Dark Matter}
\author[Mo \& Mao]
{
H. J. Mo$^1$\thanks{e-mail: hom@mpa-garching.mpg.de}
and Shude Mao$^2$\thanks{e-mail: smao@jb.man.ac.uk} \\
$^1$Max-Planck-Institute f\"ur Astrophysik,
        Karl-Schwarzschild-Strasse 1, 85740 Garching, Germany \\
$^2$University of Manchester, Jodrell Bank Observatory,
  Macclesfield, Cheshire SK11 9DL, UK}
\date{Accepted ........
      Received .......;
      in original form .......}
\pubyear{2000}

\begin{document}
\maketitle
\begin{abstract}

 We show that the Tully-Fisher relation observed
for spiral galaxies can be explained in the current scenario of galaxy 
formation without invoking subtle assumptions, 
provided that galactic-sized dark haloes have low
concentrations which do not change significantly with  
halo circular velocity. This conclusion does not depend
significantly on whether haloes have cuspy or flat
profiles in the inner region. In such a system, both 
the disk and the halo may contribute significantly to the maximum 
rotation of the disk, and the gravitational interaction 
between the disk and halo components leads to a tight 
relation between the disk mass and maximum rotation velocity.
The model can therefore be tested by studying 
the Tully-Fisher zero points for galaxies with different
disk mass-to-light ratios. With model parameters 
(such as the ratio between disk and halo mass, the specific 
angular momentum of disk material, disk formation time) chosen in 
plausible ranges, the model can well accommodate the zero-point, 
slope, and scatter of the observed Tully-Fisher relation, as well as
the observed large range of disk surface densities and sizes.
In particular, the model predicts that low surface-brightness 
disk galaxies obey a Tully-Fisher relation very similar to 
that of normal disks, if the disk mass-to-light ratio is properly
taken into account. About half of the gravitational force at 
maximum rotation comes from the disk component for normal
disks, while the disk contribution is lower for galaxies with a lower 
surface density. 

 The halo profile required by the Tully-Fisher relation
is as concentrated as that required by the observed rotation curves
of faint disks, but less concentrated than that given by
current simulations of CDM models. We discuss the implication of 
such profiles for structure formation in the universe and for
the properties of dark matter. Our results 
cannot be explained by some of the recent proposals for 
resolving the conflict between conventional CDM models and the 
observed rotation-curve shapes of faint galaxies.
If dark matter self-interaction (either scattering or
annihilation) is responsible for the shallow profile, 
the observed Tully-Fisher relation requires  
the interaction cross section $\sigma_X$ to satisfy 
$\langle\sigma_{X}\vert v\vert\rangle/m_{X}
\sim 10^{-16} {\rm cm^{3}\,s^{-1}\,GeV^{-1}}$,
where $m_X$ is the mass of a dark matter particle. 
\end{abstract}


\section {INTRODUCTION}

  It has been known since the work of Tully and Fisher (1977)
that there are systematic correlations between the luminosity and the
rotation velocity of disk galaxies. Since the rotation curves
of relatively bright disk galaxies are quite flat at large radii, 
a characteristic rotation velocity $V_{\rm c}$ (e.g. the maximum 
rotation velocity $V_{\rm max}$) can be defined
for each disk. When luminosity $L$ is plotted against
$V_{\rm max}$ for a sample of spiral galaxies, it is found that  
most galaxies lie close to a power-law relation (the Tully-Fisher
relation):
\beq\label{T_Fisher}
L= A V_{\rm max}^\alpha,
\eeq
where $\alpha$ is the slope, and $A$ is the `zero--point'.
The observed value of $\alpha$ ranges from about 2.5
to about 4; both $\alpha$ and $A$ may depend on the waveband
(Strauss \& Willick 1995 and references therein). In the $I$-band, 
recent determination by Giovanelli et al. (1997) gives
\beq\label{TF_observed}
\calM_I-5\log h=-21.00-7.68\left(\log W-2.5\right)\,, 
\eeq
where $W\approx 2 V_{\rm max}$ is the $21\cm$ hydrogen line width,
and the corresponding Tully-Fisher slope is 
$\alpha=7.68/2.5\approx 3.1$.
The observed Tully-Fisher relation is quite tight: 
for a fixed $V_{\rm max}$ the spread in the absolute
magnitude is smaller than 1 magnitude. It is possible that
the small scatter is caused by observational errors and that
the intrinsic scatter could be even smaller. Because of its tightness, 
the Tully-Fisher relation can be used as a distance indicator 
to spiral galaxies. But the tightness also gives stringent 
constraints on the formation and evolution of disk galaxies.
 Since the typical rotation velocity (such as $V_{\rm max}$) 
is related to the total gravitational mass, while $L$
is related to the total number of stars, the position of a particular 
galaxy in the $L$--$V_{\rm c}$ plane depends both on the 
efficiency for gas to be converted into stars
and on the mass distribution in the galaxy.

Real galaxy disks are observed to cover two orders of magnitude in 
luminosity and in surface luminosity density, and with a variety 
of rotation curves. The question arises whether the observed 
Tully-Fisher relation requires the conspiracy of some very 
special initial conditions for disk formation or does it have a more 
generic origin. Whatever the origin is, a successful model should
explain both the tight Tully-Fisher relation and the diversity 
of the disk population. 

 The current standard scenario of disk formation assumes
that galaxy disks form as gas cools in dark matter haloes
(e.g. Fall \& Efstathiou 1980; Dalcanton, Spergel \& Summers
1997; Mo, Mao \& White 1998, hereafter MMW).
In this scenario, the mass of a disk is determined by 
the mass of its host halo together with the efficiency
with which the halo gas settles into a disk; 
the size of a disk is determined by the angular momentum of
the disk material initially acquired from the tidal 
field of the cosmic density field, together with the 
processes that can change the disk angular momentum;  
and the rotation curve of a disk is determined by the
density profile of its host halo and the interaction 
between the halo and disk components. Detailed modelling 
shows that this scenario of disk formation, combined 
with current theory of cosmogony, is quite powerful in 
interpreting observational data on galaxy disks 
(e.g. Dalcanton et al. 1997, MMW, van den Bosch 1998, 2000;
Avila-Reese, Firmani \& Hernandez 1998; Mao, Mo \& White
1998; Mao \& Mo 1998; Heavens \& Jimenez 1999; Mo, Mao \& White 1999;
Syer, Mao \& Mo 1999; Firmani \& Avila-Reese 2000). In particular, 
MMW showed that disk formation in current CDM models can explain many
observations on galaxy disks, provided that the following conditions are 
fulfilled: 
(1) dark haloes are not very concentrated;
(2) only part of the gas in a dark halo settles into a disk;
(3) disk material does not lose much of its angular momentum 
    when it settles into a disk; 
(4) present disks form quite late, at redshifts $z\la 1$. 

Numerical simulations have also been used to study disk
formation in CDM models (e.g. Weil, Eke \& Efstathiou 1998; 
Sommer-Larsen, Gelato \& Vedel 1999; Koda, Sofue \& Wada 1999;
Navarro \& Steinmetz 2000b). One might hope that such simulations
can be used to check the assumptions made in the standard model. 
However, there is an outstanding problem at present. 
In many of these simulations, it is found 
that protogalactic gas is mostly in dense clumps.
As such clumps sink towards the halo centre to make
a disk, they lose much of their angular momentum to the
dark halo due to dynamical friction, and the resulting
disk is much too small to match any realistic galaxy disk.
One possibility to solve this problem is to assume 
that some feedback effects can keep the protogalactic gas
in a diffused form so that the effect of dynamical friction
is reduced. Such a process has not been very successfully 
implemented in current models of disk formation, but is
required in order to make progress in the standard 
framework. Recent high-resolution N-body simulations 
of dark haloes (Moore et al. 1999a,b; Jing \& Suto 2000;
Springel et al. 2000) also expose disk formation in the CDM models
to another problem: CDM haloes may be too concentrated 
to match the observed rotation curves of disk galaxies
(e.g., Navarro 1998). This mismatch  
suggests that some modifications on the CDM cosmogony might be 
needed.

In this paper we take the attitude that 
the main idea of the standard scenario for disk formation
is correct but some modifications on earlier models 
are required. As discussed above, disk formation in this
scenario involves the following important aspects: 
(1) density profiles of dark haloes, (2) the fraction of 
halo gas that can settle into a disk, (3) the specific angular
momentum of disk material relative to halo material,
and (4) the assembly times of present disks.  
We will take an empirical approach towards all these different 
aspects. Our goal is to find a simple model within the 
standard scenario, so that the most important properties
of the disk population, such as the Tully-Fisher relation,  
can be explained without invoking subtle assumptions.  
The formalism to be used is very similar to that in MMW, 
but with substantially relaxed model assumptions. 

 One of our main conclusions is that the Tully-Fisher relation 
is nothing mysterious but a generic result of the 
gravitational interaction between the disk and the halo in 
systems with properties close to that predicted by the current model.
Although this conclusion has already been reached in MMW and
Dalcanton et al. (1997), here we examine the process in 
more detail. Another main result of this paper is that the halo 
profile required for explaining the Tully-Fisher relation is 
less concentrated than that predicted by conventional CDM models 
and that the required concentration of galactic haloes does 
not depend strongly on halo mass.  
This gives strong constraints on the properties of dark matter.
In fact, our results cannot be explained by some of the recent 
proposals for resolving the conflict between conventional CDM models 
and the halo profile of faint galaxies. Assuming that
the shallow profile is due to dark matter self-interaction 
(either scattering or annihilation), we discuss 
the implication of our results for the mass and cross section 
of the dark matter particles.

\section {Simple Consideration}

\subsection {Self--Gravitating Disks}

 To start with, let us consider a simple model where galaxy disks
are self-gravitating and have exponential surface-density 
profiles:
\beq
\Sigma (R)=\Sigma_0 \exp\left(-R/\Rd\right)\,,
\eeq
where the central surface density $\Sigma_0$ and the disk 
scale-length $\Rd$ are related to the disk mass by
$\Md=2\pi \Sigma_0 \Rd^2$. The rotation curve of such a disk is
given by 
\beq\label{vc_disk}
V_{\rm d}^2(R)=4\pi G \Sigma_0 \Rd y^2\left[\calI_0(y)\calK_0(y)
-\calI_1(y) \calK_1(y)\right]\,,
~~~
y\equiv {R\over 2\Rd}\,,
\eeq
where $\calI_n$ and $\calK_n$ are modified Bessel functions 
of the first and second kinds (Freeman 1970).
This rotation curve peaks at $R\approx 2.16 \Rd$, with 
a maximum rotation velocity given by
$V_{\rm max}^2\approx 2.5 G \Sigma_0 \Rd$. Assuming a disk 
mass-to-light radio $\mtold\equiv \Md/L_\rmd$, this relation can be
written as
\beq\label{TF_model1}
L_\rmd 
=B \left({V_{\rm max}\over 200\kms}\right)^4
\left({I_0\over 100\Lsun\pc^{-2}}\right)^{-1}
\left({\mtold\over \Upsilon_\odot h}\right)^{-2}\,,
\eeq
where $\Upsilon_\odot\equiv \msun/\lsun$, 
$I_0\equiv \Sigma_0/\mtold$ is the disk central 
luminosity density, and
\beq
B\approx 8.5\times 10^{11} h^{-2}\lsun\,.
\eeq
Equation (\ref{TF_model1}) looks quite similar
to the observed relation (\ref{T_Fisher}). 
But the amplitude is too high compared with the observed
Tully-Fisher relation. To see this, we use
equation (\ref{TF_observed}) to obtain 
\beq
L_I (V_{\rm max}=200\kms)\approx 2.2\times 10^{10} h^{-2}{\lsunI}\,.
\eeq
The typical scale-length of normal galaxies with $V_{\rm max}=200\kms$
is about $3.5\kpch$ (Courteau 1996, 1997; de Jong 1996). 
The implied typical disk surface brightness 
is therefore $I_0\approx 300 \lsunI \pc^{-2}$.
This surface luminosity density is slightly higher 
than the median value of the Freeman disk, $\mu_B=21.7$ 
(Freeman 1970), with a colour correction $(B-I)=1.8$
(de Jong 1996). Since the observed colours of disk
galaxies are quite uniform, the stellar mass-to-light ratio
should not vary significantly among normal disks with 
similar luminosity. Based on various considerations --
stellar population synthesis (e.g. de Jong 1996),
stellar counts and kinematics in the solar neighborhood
(Kuijken \& Gilmore 1989; Gould, Bahcall \& Flynn 1996), 
and kinematics of external galaxies (Bottema 1997) -- 
the stellar mass-to-light ratio of galaxy disks in the $I$-band 
is about $2h$. Since the masses of normal disks are dominated 
by stars, we may write $\mtold\approx 2h$ in the $I$-band.
Inserting these values of $I_0$ and $\mtold$ into  
equation (\ref{TF_model1}), we see that the predicted 
luminosity at $V_{\rm max}=200\kms$ is about 3 times as
high as that observed. There are two possibilities 
to explain this factor of 3. First, it might be that 
the value of $\mtold$ is closer to $3h$ than to $2h$. 
Given the uncertainty in $\mtold$, such a moderate increase
may be allowed. In this case, the observed Tully-Fisher
zero-point would be consistent with the assumption
that galaxy disks are self-gravitating at the radii
of maximum rotation. The second possibility is that 
the value of $\mtold$ {\it is} close to $2h$, but about half
of the gravitational force responsible for the maximum 
rotation actually comes from an extra mass component, i.e.,
a dark halo. The second possibility is much more likely, 
not only because dark haloes are required to 
explain the flat rotation curves of spiral galaxies
(e.g. Rubin, Ford \& Thonnard 1980; Persic \& Salucci 1988)
and to stabilize a thin disk (e.g. Efstathiou, Lake \& Negroponte 1982),
but also because a model with disks dominating $V_{\rm max}$ 
cannot be made consistent with the small Tully-Fisher 
scatter. This can be seen from equation (\ref{TF_model1}):
if the mass-to-light ratio $\mtold$ is assumed not to change
significantly among galaxies with similar $V_{\rm max}$
(probably a good assumption given that their colours are quite 
uniform), the scatter in $L_\rmd$ is the same as that in $I_0$
for given  $V_{\rm max}$. 
The observed range in surface brightness 
is about 1.5 magnitudes for normal galaxies and is even larger
if low surface-brightness galaxies are included (see 
Bothun, Impey \& McGaugh 1997). The implied scatter is 
therefore much larger than that observed (see also
Courteau \& Rix 1999). Thus, the observed 
Tully-Fisher relation is not consistent with the assumption 
that disk rotation is dominated by disk gravity 
at the radius of maximum rotation.   

\subsection {The Effects of Dark Haloes
\label{ssec_haloeffect}}

 If a dark halo contributes significantly to $V_{\rm max}$
in a typical disk galaxy, we want to know whether the 
presence of dark haloes can also help to explain the 
zero-point and small scatter in the Tully-Fisher relation, 
given that disks have a large spread in their surface brightness.
Taking into account the gravity of a dark halo (assumed 
to be spherically symmetric), we should replace 
equation (\ref{TF_model1}) by
\beq\label{TF_model2}
L_\rmd 
=B \left({V_{\rm max}\over 200\kms}\right)^4
\left({\mtold\over \Upsilon_\odot h}\right)^{-2}
\left({I_0\over 100\Lsun\pc^{-2}}\right)^{-1}
\left({V_\rmd^2\over V_{\rm max}^2}\right)^2\,,
\eeq
where the total rotation velocity at the peak of the rotation curve
is a sum in quadrature of contributions from the disk and the halo:
\beq
V_{\rm max}^2=V_\rmd^2 +V_\rmh^2\,.
\eeq
For the halo density profiles to be considered in the following,
the rotation curves typically reach a maximum near 
$3\Rd$ where the disk contribution is also near its peak.
In what follows, we will identify the rotation velocity 
at $3R_\rmd$ to be the one in the Tully-Fisher
relation. For simplicity, we will still denote this velocity by
$V_{\rm max}$. Inserting the values of $I_0$ and $\mtold$ inferred 
in the last subsection into equation (\ref{TF_model2}),
we find that the ratio between the predicted 
and observed Tully-Fisher zero-points is 
\begin{eqnarray}\label{calF_defined}
{\cal F}(V_{\rm max}=200\kms)
&\equiv& {L_{\rm d, \,predicted} (V_{\rm max}=200\kms)
\over {L_{\rm d, \,observed} (V_{\rm max}=200\kms)}}
\nonumber\\
&\approx& 
1.0\times \left({\mtold\over 2h\Upsilon_\odot}\right)^{-2}
\left({I_0\over 300\lsunI \pc^{-2}}\right)^{-1}
\left[{(V_\rmd/V_{\rm max})^2\over 0.5}\right]^2\,.
\end{eqnarray}
We see that $V_\rmd^2/V_{\rm max}^2 \approx 0.5$ is required for 
the model prediction to match the observation.
In other words, dark haloes must contribute about half of the
gravitational force at the radius of maximum rotation for 
a typical disk galaxy with $V_{\rm max}\sim 200\kms$.

 To see if the presence of a halo also helps to reduce the 
scatter, let us consider a `typical' disk with 
mass $\Md$ and surface density $I_0$. 
As before we assume $\mtold$ to be invariant from galaxy
to galaxy, so that $\Md$ is equivalent to $L_{\rmd}$ and
$I_0$ is equivalent to $\Sigma_0$. Now suppose we increase
the disk mass from $\Md$ to $(1+\epsilon)\Md$ in such a way
that disk scale-length is fixed, i.e. we increase  the
disk mass without changing the disk concentration in the halo.
In this case, both $I_0$ and $V_{\rmd}^2$ are increased by 
a factor of $(1+\epsilon)$. Suppose such an increase in $\Md$ 
induces a change $\delta V_\rmh^2$ in the halo contribution, the ${\cal F}$ 
factor defined in equation (\ref{calF_defined}) becomes
\beq
{\cal F}' ={1+\epsilon \over (1+\epsilon')^2}{\cal F}
~~~\mbox{with}~~~~
\epsilon'\equiv {\epsilon V_{\rmd}^2 +\delta V_\rmh^2\over V_{\rm
max}^2}\,.
\eeq
If the maximum rotation is dominated by the disk, then 
$\epsilon'\approx \epsilon$ and 
${\cal F}'\approx {\cal F}/(1+\epsilon)$, i.e. the Tully-Fisher
zero-point is {\it reduced} by a factor of $(1+\epsilon)$. 
In this case, we recover the result for self-gravitating disks.
If, on the other hand, disk gravity is negligible, then
$\epsilon'\approx 0$ and 
${\cal F}'\approx {\cal F}(1+\epsilon)$, i.e. the Tully-Fisher
zero-point is {\it enhanced} by a factor of $(1+\epsilon)$. 
In this halo-dominated case, the disk rotation curve is independent 
of the disk mass, and so the Tully-Fisher zero-point is
directly proportional to the disk mass. 
Thus, somewhere between the disk dominating and the halo dominating,
the Tully-Fisher zero-point will not be altered by the change
of $\Md$. In fact, if both $\epsilon$ and $\epsilon'$ are small, 
and if $\delta V_\rmh^2$ is neglected, then 
${\cal F}'={\cal F}$ for $(V_\rmd/V_{\rm max})^2\approx 0.5$,
i.e. the Tully-Fisher zero-point does not depend on $\Md$
if the disk contributes about half of the gravitational force 
at the maximum rotation. As another example, if we increase 
$I_0$ (or $\Sigma_0$) by a factor of $(1+\epsilon)$ 
but keep $\Md$ unchanged, i.e. if we increase the disk 
concentration in the halo without changing the mass,
then we have for small $\epsilon$,
\beq\label{calF2}
{\cal F}' ={1\over (1+\epsilon')^2}{\cal F}
~~~\mbox{where}~~~~
\epsilon'\equiv 
{(\epsilon/2) V_{\rmd}^2 +\delta V_\rmh^2\over V_{\rm
max}^2}\,.
\eeq
So, ${\cal F}'\approx {\cal F}$ for
halo-dominating and ${\cal F}'\approx {\cal F}/(1+\epsilon)$
for disk-dominating. The above simple arguments suggest 
that the presence of dark haloes can help to 
reduce the Tully-Fisher scatter. However, the details 
depend on $\delta V_\rmh^2$, the halo response to the 
change in the disk, which we will study in the following section.

\section {Detailed Modelling}

 In order to examine in more detail the constraints
provided by the Tully-Fisher relation on disk formation,
we consider exponential disks in realistic dark haloes. 
Our modelling here follows the disk formation model of MMW.
The reader is referred to that paper
for details; here we only repeat the essentials of the model.
Briefly, after the initial protogalactic collapse the gas and dark
matter are assumed to be uniformly mixed in a virialized object.
As a result of dissipative and radiative processes,
the gas component gradually settles into a disk. 
We assume that the mass of this disk is a fraction $\md$  
of the halo mass, and that its specific angular momentum is 
$\jd$ times that of the dark halo. 
If the mass profile of the disk is taken to be exponential, 
and if the dark halo responds to the growth of the disk adiabatically
(see Barnes \& White 1984 and Navarro \& Steinmetz 2000b
for tests of the validity of this assumption), 
then $\Sigma_0$, $\Rd$ and the galaxy's rotation curve
are determined uniquely. Specifically,
\beq\label{rd_sis}
\Rd = {1\over \sqrt{2}} \left({\jd\over\md}\right)
\lambda r_{\rm h} F_R\,,
~~~
\Sigma_0 = {\md M_\rmh \over 2 \pi \Rd^2}\,,
~~~
V_{\rm max}= V_\rmh F_V\,,
\eeq
where $\lambda$ is the spin parameter of the halo,
$M_\rmh$, $V_\rmh$ and $r_\rmh$ are the mass, circular velocity 
and virial radius of the halo before responding to 
disk gravity, $F_R$ and $F_V$ are factors which depend
on the halo profile and disk action. For a given 
cosmology, $M_\rmh$, $V_\rmh$ and $r_\rmh$ are 
related by
\beq\label{Mhasz}
r_\rmh={V_\rmh  \over 10 H(z)}\,,~~~~
M_\rmh={V_\rmh^3\over 10 G H(z)}\,,
\eeq
where $H(z)$ is the Hubble constant at redshift $z$
(see MMW for details).

 Let us consider a case where the halo density profiles
have the form 
\begin{eqnarray}\label{profile}
\rho (r)&=&\rho_0 {r_{\rm c}^3\over (r+r_{\rm c})^3}\nonumber\\ 
        &=& {V_{\rm h}^2\over 4\pi G r^2}
{c\over \left[\ln(1+c)-c(1+3c/2)/(1+c)^2\right]}
{r^2/r_{\rm c}^2\over (r/r_{\rm c}+1)^3}\,,
\end{eqnarray}
where $r_{\rm c}=r_{\rm h}/c$ is a core radius and
the quantity $c$ is known as the halo concentration factor. 
The above profile can be compared with the NFW profile:
\begin{eqnarray}\label{NFW}
\rho (r)&=&\rho_0 {r_{\rm s}^3\over r (r+r_{\rm s})^2}\nonumber\\ 
&=&{V_{\rm h}^2\over 4\pi G r^2}
{c_{\rm NFW}\over \left[\ln(1+c_{\rm NFW})
-c_{\rm NFW}/(1+c_{\rm NFW})\right]}
{r/r_{\rm s}\over (r/r_{\rm s}+1)^2}\,,
\end{eqnarray}
where $r_\rms= r_{\rm h}/c_{\rm NFW}$ is a scale radius
(Navarro, Frenk \& White, 1997, hereafter NFW), and 
$c_{\rm NFW}\equiv r_{\rm h}/r_{\rm s}$ describes the halo 
concentration in this profile. Although profiles (\ref{profile})
and (\ref{NFW}) behave differently at small radii, 
the two can be made to have similar global properties 
by properly choosing the values of $c$ and $c_{\rm NFW}$.
For example, galactic-sized haloes with $V_\rmh\sim 200\kms$
in CDM models have $c_{\rm NFW}\approx 20$ (Moore et al. 1999b;
see also the value quoted in Navarro \& Steinmetz 2000a). 
The corresponding profile looks
similar to profile (\ref{profile}) with $c=40$ over a large range 
of radii. Since our results depend mainly on the halo 
concentration and only weakly on the details of the 
halo profile, we will present our results based on
profile (\ref{profile}) with $c$ as a free parameter, 
but we will also refer to the results obtained for the NFW profile. 

Disk formation in these kind of dark haloes is described 
in considerable detail in MMW, and the procedure outlined there
can be used to calculate $F_V$ and $F_R$ as functions of $c$, 
$j_\rmd\lambda$ and $\md$ for a given profile.  From 
these we can calculate the affects of changing 
model parameters on the Tully-Fisher relation and  the
disk size (or surface density).
\begin{figure}
\vskip-0.6cm
\centering\leavevmode
\epsfxsize=0.85\columnwidth\epsfbox{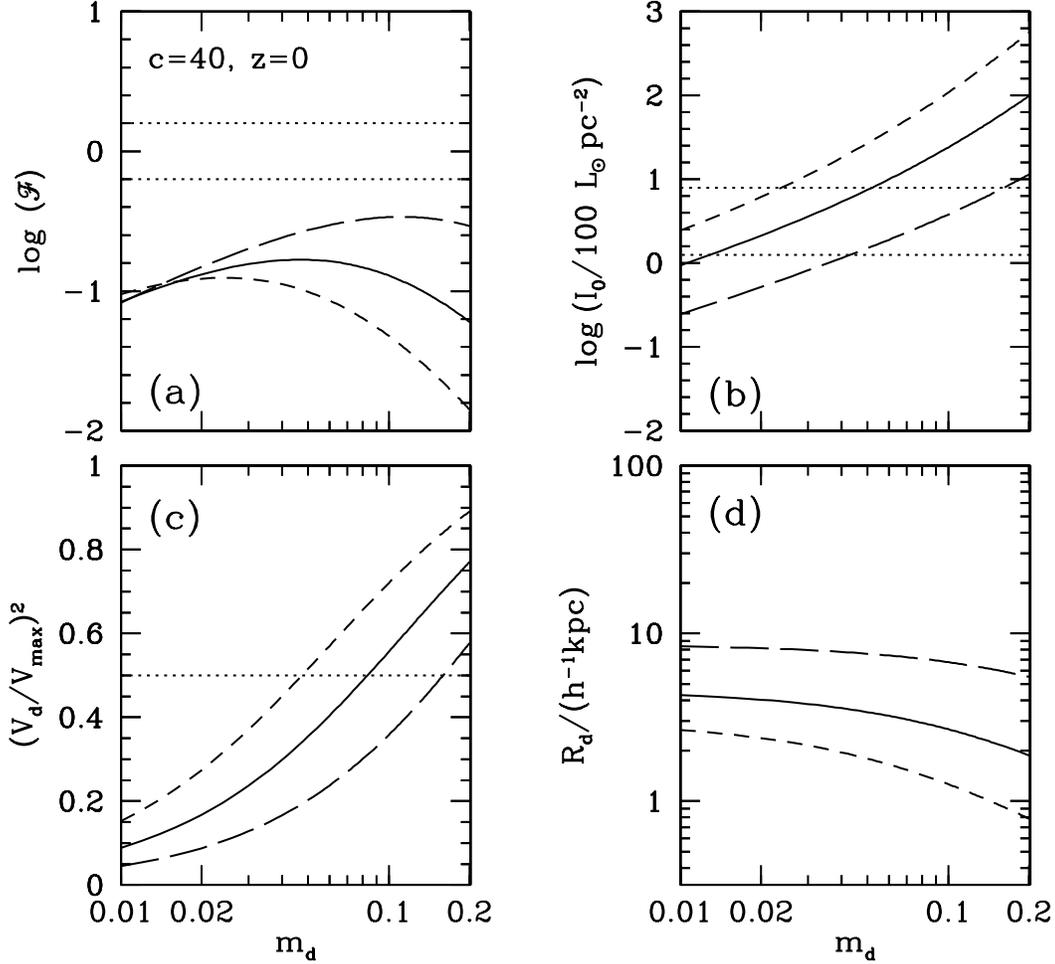}
\vskip-2.0cm
\caption{ (a) The change in the Tully-Fisher zero-point
for haloes with $V_\rmh=200\kms$ and $c=40$.
[Notice that profile (\ref{profile}) with $c=40$ matches the NFW profile
with $c_{\rm NFW}\approx 20$.]
Curves from top down have $\lambda=0.1$ (long-dashed curve), 
$0.05$ (solid curve) and $0.03$ (short-dashed).
The two horizontal lines bracket the range of ${\cal F}$ 
required to match the observed Tully-Fisher zero-points among 
different galaxies at $V_{\rm max}\sim 200\kms$.
Panel (b) is the predicted central luminosity density.
Curves from bottom up have $\lambda=0.1$, $0.05$ and $0.03$.
The two horizontal lines indicate roughly the observed range of 
$I_0$ in the $I$-band for normal galaxies 
(the median value is taken to be $I_0=300 \lsun\pc^{-2}$
and the range of surface brightness is taken to be 
$\pm 1$ magnitude to represent approximately 
the observed range of Freeman disks).  
Panel (c) shows the disk contribution to the total rotation.
Curves from bottom up have $\lambda=0.1$, $0.05$, and $0.03$.
The horizontal line separate disk-dominating from halo
dominating. Panel (d) shows the disk scale-length.
Curves from bottom up have $\lambda=0.03$, $0.05$, and $0.1$.}
\label{fig_fig1}
\end{figure}
\begin{figure}
\vskip-0.6cm
\centering\leavevmode
\epsfxsize=0.85\columnwidth\epsfbox{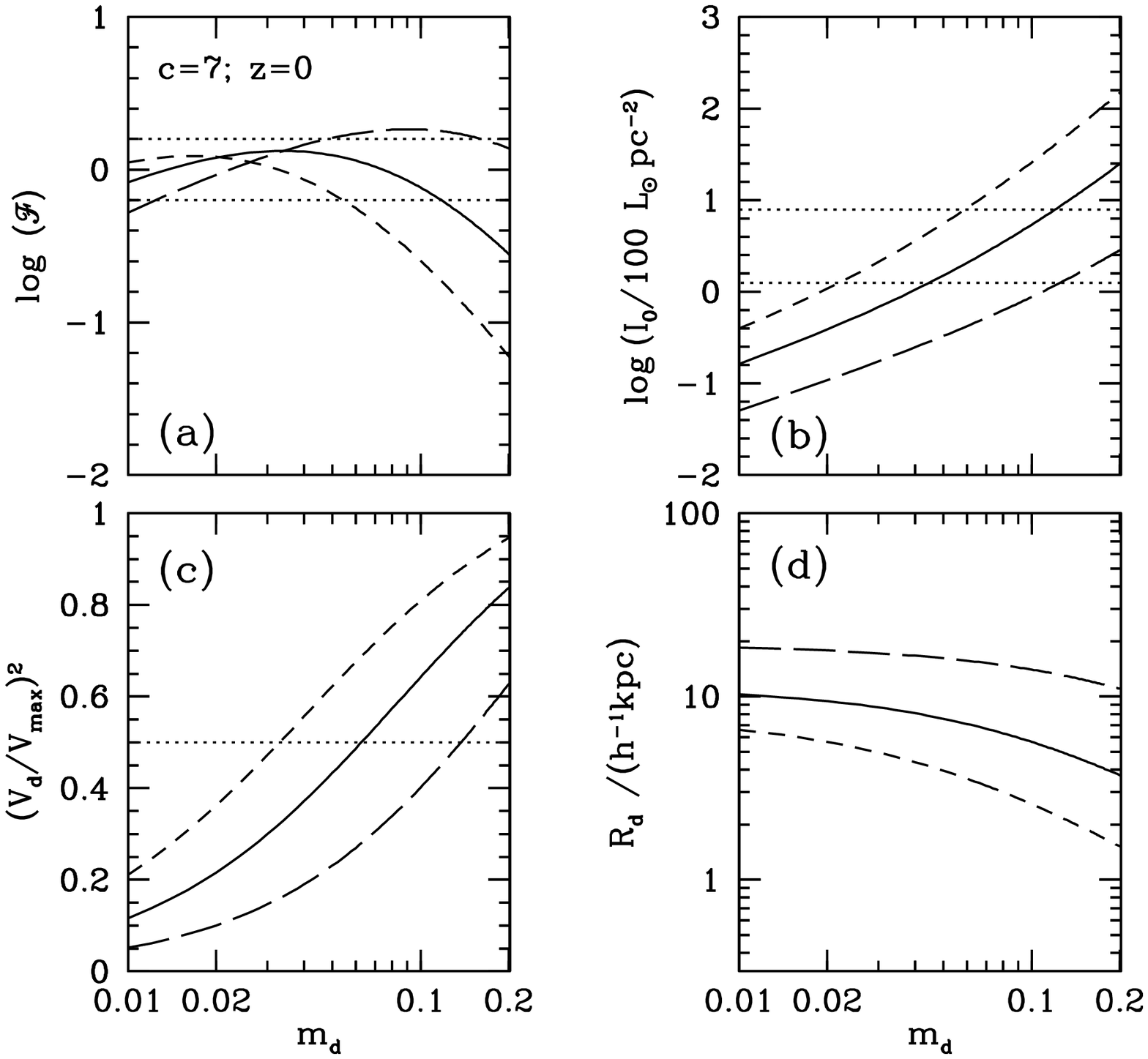}
\vskip-2.0cm
\caption{The same as Fig.\,\ref{fig_fig1},
but the concentration factor $c$ is assumed to be 7.
Profile (\ref{profile}) with $c=7$ matches the NFW profile
with $c_{\rm NFW}\sim 3$.}
\label{fig_fig2}
\end{figure}
Figures \ref{fig_fig1} and \ref{fig_fig2} show the results 
of such calculations. Several important conclusions can be reached 
from these results. As the value of $\md$ changes from 0.01
to 0.1, the central surface density $I_0$ changes 
by a factor of about 30, but the Tully-Fisher 
zero-point [described by the factor ${\cal F}$ defined in
equation (\ref{calF_defined})] changes only by a factor 
of about 3 for any given $\lambda$. The value of $m_\rmd$ must be
smaller than the overall baryon fraction in the universe,
$\Omega_{\rm B,0}/\Omega_0$ (which is, according to cosmic
nucleosynthesis, about $0.05$ for an Einstein-de Sitter universe
with $h=0.5$, and about $0.1$ for a low-density universe
with $\Omega_0=0.3$ and $h=0.7$). Also $m_\rmd$ should not be 
much smaller than 0.01, because such disks, if they exist, 
may not be able to form enough stars [due to the failure to meet the 
Toorme (1964) instability criterion] to be included in 
current Tully-Fisher samples. Therefore, any reasonable 
change in $\md$ will not introduce a large Tully-Fisher scatter.
This is somewhat contrary to intuition. If disk gravity were
negligible, an increase of $m_\rmd$ by a factor of $10$
would increase the Tully-Fisher zero-point by a factor 
of about 10; if, on the other hand, halo gravity were negligible, 
an increase of $m_\rmd$ by a factor of 10
would decrease the Tully-Fisher zero-point by a factor 
similar to that in $I_0$, i.e. of about 30. The reason for the 
insensitivity of the Tully-Fisher zero-point to the change of $m_\rmd$ is
that the interaction between the disk and the halo acts to reduce the 
scatter, as is demonstrated in Section \ref{ssec_haloeffect} by
simple analytic arguments. Indeed, the 
Tully-Fisher zero-point is almost independent of $\md$ in the range where 
$V_{\rmd}^2\sim V_{\rm max}^2/2$; it increases
with $\md$ at $V_{\rmd}^2 < V_{\rm max}^2/2$ and decreases with
$\md$ at $V_{\rmd}^2 > V_{\rm max}^2/2$.
This behaviour is exactly what is expected from the simple 
model discussed in Section \ref{ssec_haloeffect}.
Similar effect is found in the simulations of  
Navarro \& Steinmetz (2000b). In their simulations, Navarro \& Steinmetz 
defined a Tully-Fisher relation using disk rotation velocities
measured at large radii which are proportional to the halo
circular velocities. Because of the difference in definition,
it is not straightforward to relate our results to their
simulation results.

\begin{figure}
\vskip-0.6cm
\centering\leavevmode
\epsfxsize=0.85\columnwidth\epsfbox{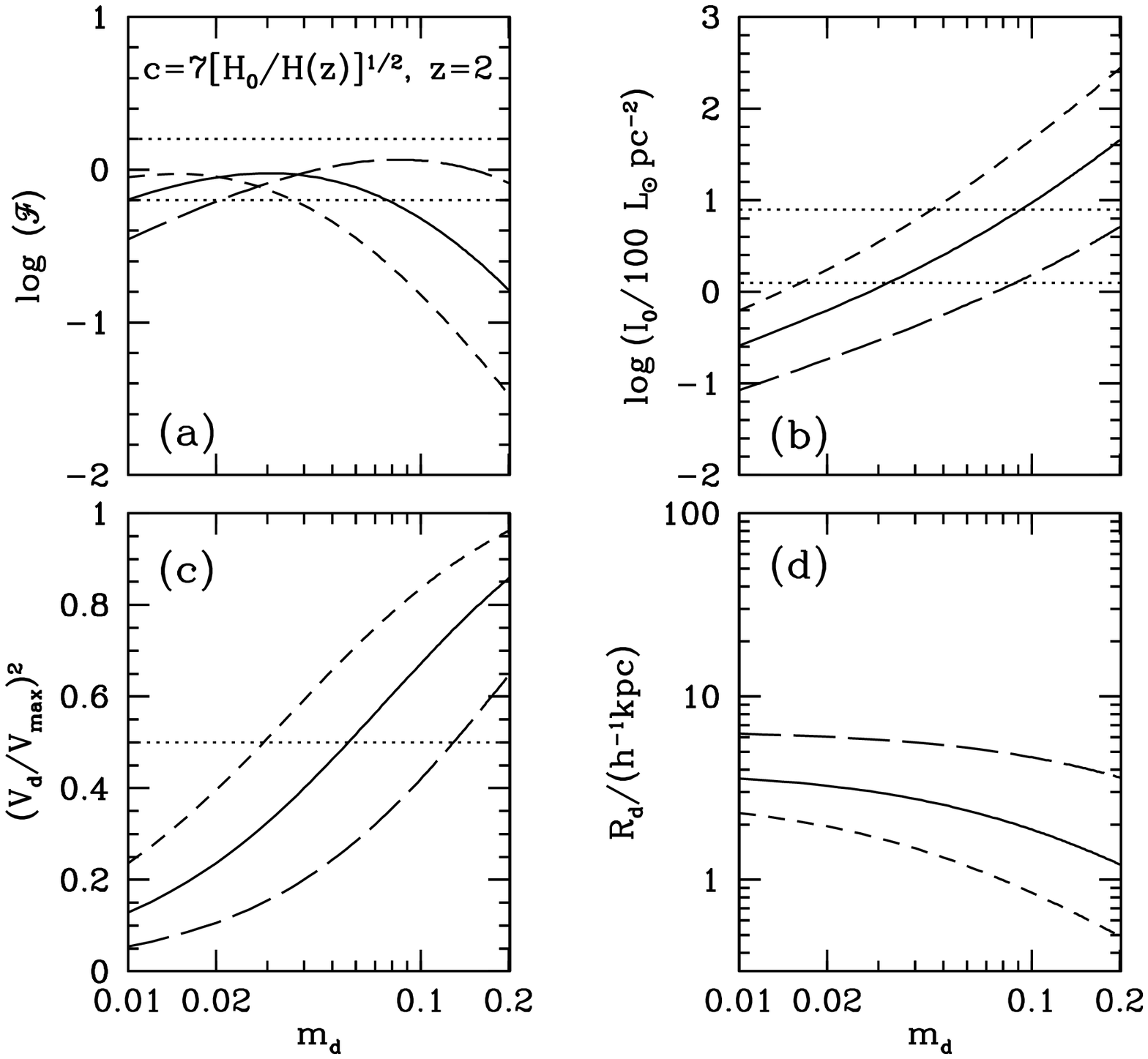}
\vskip-2.0cm
\caption{The same as Figure \ref{fig_fig1}, except that
disks here are assumed to be assembled at $z=2$ in haloes
with $V_{\rm h}=200\kms$ and $c=7[H_0/H(z=2)]^{1/2}$. 
A low-density cosmology, with $\Omega_0=0.3$ and $\Lambda=0.7$,
is used in the calculation.
Notice that the curves are very similar to those shown 
in Figure \ref{fig_fig2}, except that disk scale-length 
is smaller by a factor of about $H(z=2)/H_0$.}
\label{fig_zeffect}
\end{figure}

 The distribution in the spin parameter $\lambda$ also introduces
scatters in the Tully-Fisher zero-point.
Since changing $\lambda$ is equivalent to changing the disk
concentration in a halo, the effect is smaller for
smaller $m_\rmd$, as implied in equation (\ref{calF2}).
However, the scatter induced by the $\lambda$-distribution is not 
expected to be large. From N-body simulations we know that 
the spin parameters of dark haloes have log-normal distribution 
with a median value ${\overline\lambda}\approx 0.05$ and dispersion
$\sigma_{\ln\lambda}\approx 0.5$ (see eq. [15] in MMW). If disks have similar
specific angular momenta as their dark haloes, then 
only 10 percent of the systems have $\lambda\ge 0.1$ and
10 percent of them have $\lambda\le 0.025$. Such a spread
in $\lambda$ does not induce too  large a scatter in the 
Tully-Fisher relation, as shown in MMW and as can also be 
inspected from Figures \ref{fig_fig1} and \ref{fig_fig2}.
However, if disks could lose a large fraction of their angular momenta so that
their effective spins were much lower than 0.05, 
they would be too compact and the Tully-Fisher zero-point 
would be too low. 
Notice that disks with large $m_{\rm d}$ and small $\lambda$,
which are predicted to have systematically low Tully-Fisher 
zero-point, may be unstable, and so the scatter induced by
the $\lambda$-distribution may be reduced if such systems
do not form real disks. But significant loss of
angular momentum would then lead to too few stable disks.
Another important result shown in 
Figure \ref{fig_fig1} and \ref{fig_fig2} is that low surface-brightness disks
(formed in systems with high $\lambda$ and low $m_\rmd$)
 have a Tully-Fisher zero-point similar to that of 
`normal' disks with a surface-brightness close to that
of a Freeman disk. Thus, the observational fact that 
low surface-brightness galaxies obey a Tully-Fisher 
relation similar to that of normal spiral galaxies (Zwaan et al. 1995,
but also see O'Neil, Bothun, \& Schombert 1999) can be 
explained here without invoking any subtle assumptions.
In reality, however, some offset in the Tully-Fisher zero-point 
is expected for the low surface-brightness population.
Some low surface-brightness disks may contain more cold gas
(because of low star formation efficiency)
than normal galaxies (e.g. McGaugh \& de Blok 1997;
O'Neil, Bothun, \& Schombert 1999), and so their Tully-Fisher 
zero-point is expected to be lower because of their higher 
disk mass-to-light ratios. Our results suggest that
low-surface brightness galaxies should obey the same
Tully-Fisher relation as high surface-brightness galaxies, 
if the variation of the gas fraction is properly taken into account.
This is in agreement with the observational results
that the relation between rotation velocity and disk 
mass is tighter than the relation between rotation 
velocity and luminosity (e.g. McGaugh et al. 2000).

 The change of halo concentration $c$ has quite a large effect on the predicted
zero-point. To match with observations, the value
of $c$ is required to be low, $c\sim 7$. Similar match can be 
obtained for the NFW profile with $c_{\rm NFW}\sim 3$. 
The implied halo concentration is therefore much lower than 
the value ($c_{\rm NFW}\sim 20$) obtained from N-body simulations
of CDM models for galactic-sized haloes. Since a lower 
$c$ value means lower concentration of haloes, 
the result suggests that real galaxy
haloes must be much less concentrated than CDM haloes.
This low value of $c$ is in fact required by the 
rotation-curve shapes of galaxies with low luminosity 
and low surface brightness (e.g. Navarro 1998).
The result here suggests that haloes with the same low  
concentrations are also needed for normal galaxies.

 Another source that can cause scatter in the Tully-Fisher
relation is the redshift distribution of the disk assembly.
As one can see from equation (\ref{Mhasz}), for a given
halo circular velocity $V_\rmh$, haloes at redshift $z$
are lighter by a factor of $H(z)/H(0)$ than at $z=0$.
If the other parameters (i.e. $\md$, $\lambda$,
$c$, and $\mtold$) are kept the same, disks in haloes with the same 
$V_\rmh$ have the same $V_{\rm max}$ without depending on 
redshift, and so disks assembled at higher redshifts would have 
lower luminosity. The effect could be quite large. At $z=1$, 
$H(z)/H(0)$ is about 2.8 for an Einstein-de Sitter universe, 
and about 1.8 for a flat universe with $\Omega_0=0.3$, $\Lambda=0.7$. 
This factor becomes 5.2 and 3.0 at $z=2$ for these two cosmologies. 
Thus, unless present disks have quite uniform formation times,
the induced scatter would be too large. This problem was noticed by MMW,
and they solve it by assuming most of the present disks 
to be assembled at $z\la 1$. In this case, disk formation 
in a flat universe with $\Omega_0=0.3$, $\Lambda=0.7$
can be made compatible with the observed Tully-Fisher zero-point and
scatter\footnote{Notice that the halo concentration
used in MMW is lower than that given by recent 
high-resolution N-body simulations. The disk instability
criteria used there also act to increase the Tully-Fisher 
zero-point.}.

Here we suggest another possibility. If haloes at 
higher redshift are less concentrated, we can inspect from 
Figure \ref{fig_fig2} that the redshift effect 
on the Tully-Fisher zero-point is reduced. In fact, if the
halo concentration decreases with redshift $z$ as 
$[H(z)]^{-\beta}$, with $\beta\sim 0.5$ -- 1, 
then the redshift effect can be removed almost completely
for $z\la 2$. This is shown in Figure \ref{fig_zeffect}.
The dependence of $c$ on $z$ in the NFW model is quite weak,
but recent simulation results by Bullock et al. (1999) show that
a strong $z$-dependence of $c$ is possible.

\begin{figure}
\vskip-0.6cm
\centering\leavevmode
\epsfxsize=0.85\columnwidth\epsfbox{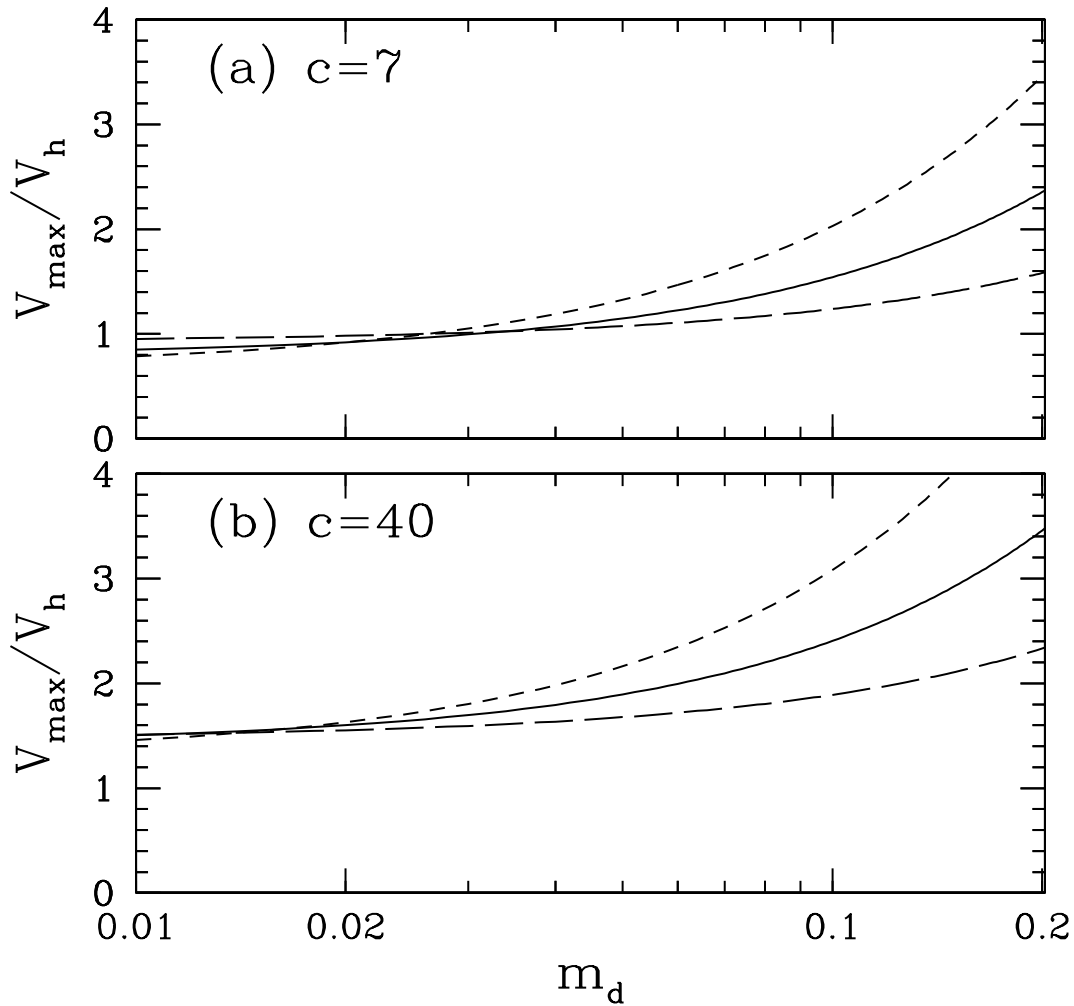}
\vskip-1.0cm
\caption{The ratio between $V_{\rm max}$ (disk rotation velocity
at $3 R_\rmd$) and the halo circular velocity $V_{\rm h}$ in models
with (a) $c=7$ and (b) $c=40$. The short-dashed, solid, and long-dashed
curves show results for $\lambda=0.03$, 0.05, and 0.1,
respectively.}
\label{fig_boost}
\end{figure}

\begin{figure}
\vskip-0.6cm
\centering\leavevmode
\epsfxsize=0.85\columnwidth\epsfbox{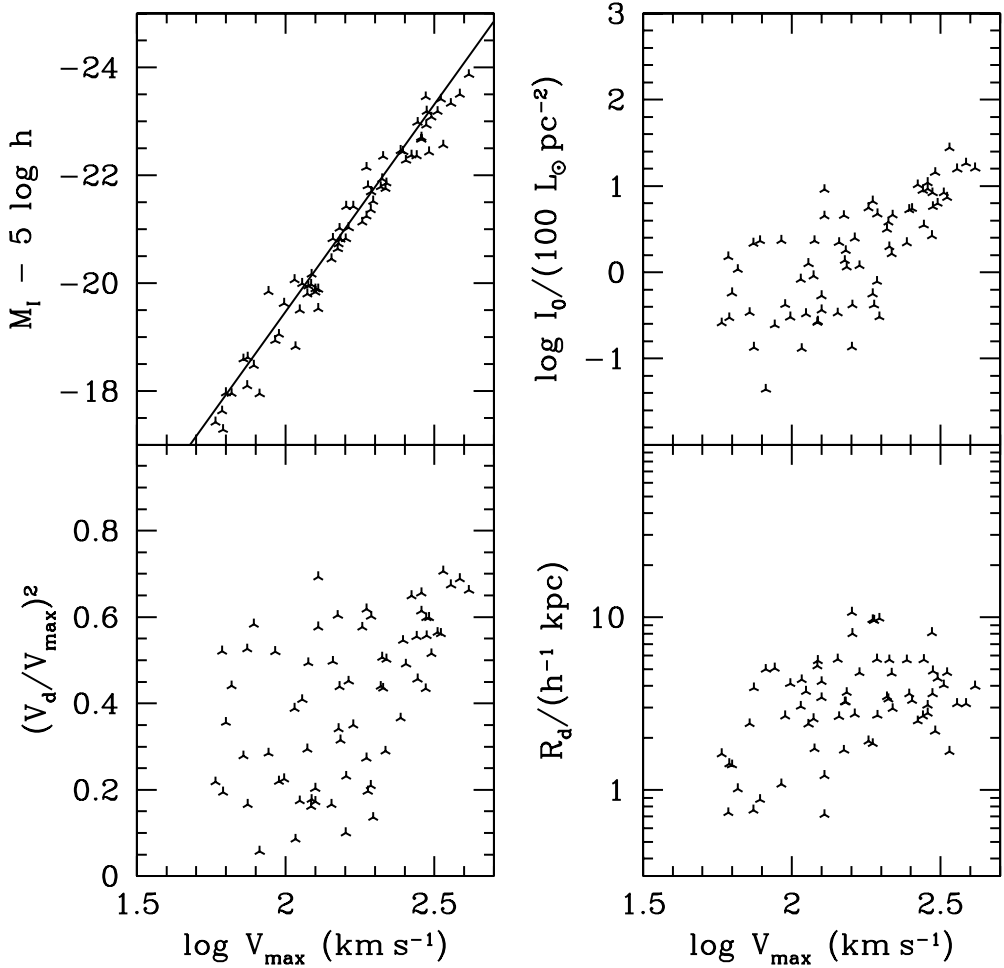}
\vskip-1.0cm
\caption{Various disk properties (obtained by Monte Carlo simulations)
plotted vs. $V_{\rm max}$ (defined to be
the circular velocity at three disk scale-lengths).
The redshift of the disk assembly is assumed to have  
uniform distribution over 0 to 2, $\md$ is assumed to have 
uniform distribution from 0.01 to 0.1, and the 
spin parameter $\lambda$ follows the log-normal distribution with a
lower-cutoff of 0.03. The concentration parameter
decreases with redshift as $7 [H_0/H(z)]^{1/2}$. 
A low-density cosmology, with $\Omega_0=0.3$ and $\Lambda=0.7$,
is used in the calculation. The top left panel shows the
Tully-Fisher relation, where the solid line indicates the observed
relation by Giovanelli et al. (1997). The top right panel shows the
central surface luminosity density. The bottom left panel shows the disk
contribution to the circular velocity at three disk scale-lengths while
the bottom right panel shows the disk scale-length vs. $V_{\rm max}$.
}
\label{fig_scatterplot}
\end{figure}

  In the model where the observed Tully-Fisher zero-point is 
reproduced, the distribution in $I_0$ is
broader than that of Freeman disks. In particular, the model
predicts the existence of low-surface brightness galaxies
in systems with high $\lambda$ and small $m_\rmd$. This is
consistent with the fact that galaxy disks with a surface brightness 
lower than that of a Freeman disk are observed in 
deep photometric observations (Bothun et al. 1997).
As one can also see from Figure \ref{fig_fig2}, 
the contribution of the disk component to the maximum 
rotation varies significantly. Generally, the maximum rotation 
velocity is dominated by the disk component in systems
with high disk surface brightness, and becomes halo dominated
in low surface-brightness systems.
 There is still intense debate whether the observed disk galaxies 
are maximal or not (e.g., Bottema 1997; Courteau \& Rix 1999; 
Debattista \& Sellwood 1998; Englmaier \& Gerhard 1999;
Bosma 2000). For our own Galaxy, the observed disk scale-length is about
$3.5\kpc$. The dark matter mass within a radius $10\kpc$ 
(which is about $3R_\rmd$) is about $5\times 10^{10}\msun$
(e.g. Dehnen \& Binney 1998). Using a rotation velocity of 
$220\kms$ at this radius, we have $(V_\rmd/V_{\rm max})^2\sim 0.5$,
which is
in good agreement with our prediction (see
the bottom left panel in Figure \ref{fig_scatterplot}.)
If the dark halo of the Milky Way were as concentrated as that predicted 
by CDM models, then the predicted value of $(V_\rmd/V_{\rm max})^2$
would be much smaller. Navarro \& Steinmetz (2000a) have 
used this to argue against CDM models. 

 Figure \ref{fig_boost} shows the ratio between $V_{\rm max}$ and the
halo circular velocity $V_\rmh$. As one can see, the boost 
in the velocity is substantial in systems with high $c$, 
high $m_\rmd$ and low $\lambda$. For $c=7$, significant 
boost occurs for $m_\rmd>0.05$ and $\lambda <0.03$, while for
$c=40$ the boost is significant for all values of $m_\rmd$ and 
$\lambda$ because of the concentrated halo profile.

 Our discussion so far has been based on systems with a
halo circular velocity $V_{\rm h} =200\kms$. Since the model 
described above also reproduced the Tully-Fisher slope
(see MMW), the discussion is also valid for other 
$V_\rmh$. As a summary, we show in Figure \ref{fig_scatterplot}
disk properties versus $V_{\rm max}$ for a Monte-Carlo 
sample, where the halo circular velocity changes uniformly between
50 and 250$\,{\rm km\,s^{-1}}$, $m_\rmd$ has a uniform distribution 
from 0.01 to 0.1, $\lambda$ has the log-normal distribution 
discussed above but with a lower cutoff at 0.03
(the small number of systems with $\lambda<0.03$ may produce 
disks that are too compact to be globally stable, see MMW
for a discussion), disk assembly redshift has 
uniform distribution between $z=0$ and $z=2$, and the
halo concentration $c$ changes with redshift as $c=7[H_0/H(z)]^{1/2}$.  
 From Figure \ref{fig_scatterplot}, it is clear that 
although $m_{\rm d}$ and formation redshift
are allowed to vary in substantial ranges, the scatter around the
Tully-Fisher relation is still compatible with observations. The
predicted central luminosity density (top right)
and disk sizes (bottom right) are also consistent
with observations (Courteau 1996, 1997; de Jong 1996).
The bottom left panel in Figure \ref{fig_scatterplot} shows the disk
contribution to $V_{\rm max}$; it is quite clear that
some systems are disk dominated while others are not. 
Disk domination of $V_{\rm max}$ preferentially occurs in systems 
with high $\md$ and low $\lambda$.

 As mentioned above, although the results presented here are 
based on the special functional form (\ref{profile}) for the halo density 
profile, our conclusions are not altered if other  
reasonable forms are used. This is not surprising, because
the quantities we are interested in here are global 
properties of the halo/disk systems. Thus, the most important 
requirement is that dark haloes have low concentrations, while the exact 
form of the halo profile is not stringently constrained 
by the global properties considered here. 
Detailed modelling of disk rotation curves is needed 
in order to see which model fares better. 
 
\section {Discussion}

In this paper, we have applied the same formalism as in MMW, but with
substantially relaxed assumptions, to study the properties of disk
galaxies. We find that even if we allow the disk mass and formation
redshift to vary substantially, the observed properties (including the
Tully-Fisher relation, disk sizes and central luminosities) can still be
reproduced. The Tully-Fisher relation is a generic 
result of the gravitational interaction between the disk and halo
components in a disk/halo system with properties close to
those predicted by current models. The model prediction
is a relation between the disk mass and the maximum rotation velocity
(the Tully-Fisher relation of mass). 
For a given halo concentration, this relation can be written
$M_{\rm d} \propto V_{\rm max}^\alpha$ with $\alpha\approx 3$.
If the halo concentration changes systematically with halo
circular velocity, the value of $\alpha$ will be different from
$3$. For example, if haloes with smaller circular velocities 
are systematically more concentrated, then the zero-point 
of the mass Tully-Fisher relation will decrease with decreasing halo 
circular velocity, leading to a higher value of $\alpha$.
To predict the Tully-Fisher relation of light, we need to know the 
mass-to-light ratios of disks. Clearly, disks with higher 
mass-to-light ratios are predicted to have lower Tully-Fisher zero-points. 
Thus, if the disk mass-to-light ratio varies systematically with halo 
circular velocity,  the slope of the Tully-Fisher relation 
for the light is expected to be different from that for the mass. 
These predictions can be tested by analysing the Tully-Fisher zero points for 
galaxies with different disk mass-to-light ratios. 

 In order to reproduce the observed Tully-Fisher relation 
in the $I$ band, we have found that the galactic haloes 
must have quite low concentrations. 
The requirement of low halo concentration is consistent with direct 
modelling of rotation curves, particularly for low surface-brightness
galaxies; the low central density of dark matter can also explain why
bars in galaxies seem to rotate rapidly (Debattista \& Sellwood 1998).
The required concentration is, however, lower than that
found in numerical simulations (NFW; Jing \& Suto 1999; 
Moore et al. 1999b; Springel et al. 2000). 

 The low halo concentration required by disk galaxies 
has many important observational consequences. For example, 
the lensing properties of disk galaxies may be different from 
those in models where concentrated profiles are assumed
(e.g. Bartelmann \& Loeb 1998). There are already a number of lenses
that appear to be caused by disk galaxies, 
such as 2237+0305 (Huchra et al. 1984), 1600+4344 (Jackson et al. 1995) and 
0218+357 (Patnaik et al. 1993). None of these systems shows a central
image, which implies that the core radii for these lensing
disks may be fairly small. It would be very interesting to use 
these systems to put quantitative constraints on the core radius. 
Similarly, the lack of central images in elliptical lenses
also suggests that the total density profile (baryons plus dark
matter) is near singular in the central region, i.e. the core radius 
must be small (e.g. Kochanek 1996). 
An important question is whether the halo profiles
of elliptical galaxies have similar core radii as disk galaxies.
It is possible that the dark halo profiles in elliptical galaxies
are significantly modified by dynamical processes during formation 
(e.g., by merging).

 Recent high-resolution N-body simulations of the formation
of cold dark matter haloes show that such haloes generally 
contain too many subclumps to match the number of dwarf 
galaxies observed in the Local Group. This happens because 
the CDM particles which form a dark halo were 
generally in progenitors with high central densities. As such
progenitors merge to form a larger halo, their central parts
can survive as subclumps (Moore et al. 1999a; Klypin et al. 1999;
Springel et al. 2000). However, if the progenitors have 
lower concentrations, they are more likely 
to be destroyed by the merging process, and the number 
of subclumps in dark haloes may be reduced. Also, 
if dark haloes have flat central cores and if the core radius 
of a halo does not change much with redshifts, we would expect 
a limiting redshift beyond which dark haloes may not be able
to form in large number. This may help to alleviate the overcooling 
problem in CDM models where too much gas can cool in small
haloes at high redshift, but is constrained by 
the fact that large numbers of star forming galaxies 
are observed at redshift $z\ga 3$ (e.g. Steidel et al. 1998).    
Clearly, many of these issues need thorough investigation
before any definite conclusions can be drawn.

  An equally important issue is the origin of the required
profiles. Since highly-concentrated 
profiles are quite generic in conventional CDM 
models, the formation of haloes with low concentrations
may require some modification in the properties of the dark matter particle 
(e.g. Spergel \& Steinhardt 1999; Hogan \& Dalcanton 2000) or some
change in the power spectrum of initial perturbations 
(Kamionkowski \& Liddle 1999). 

 In the proposal of Spergel and Steinhardt, dark matter is 
assumed to be self-interacting. The initial hope was that 
the collisions between dark matter particles can heat up the 
low-entropy material and thereby produce a shallower density 
profile. However, recent numerical simulations of the formation
of collisional haloes, with dark matter treated as a fluid,
show that the inner density profiles are even steeper than their 
collisionless counterparts (Moore et al. 2000; Yoshida et al. 2000).
Collisional dark halo might also be too spherical
to match the elliptical shape of clusters like MS21137-23 as inferred
from gravitational lensing (Miralda-Escude 2000).  
But the situation is not yet clear. The simulations of Burkert 
(2000) taking into account the effect of finite collisional 
cross-section show that core-like structure {\it can} be produced. 
The question is then whether such models can produce the low 
concentrations required to explain the Tully-Fisher zero-point.
As we have pointed out before, in order to match the observed
Tully-Fisher relation, it does not matter much whether we can get
rid of the central cusps, it is the global halo concentration 
that matters. In the proposal 
of Hogan and Dalcanton, galactic haloes are assumed to be dominated 
by warm dark matter with initial velocity dispersion. In this case,
the initial adiabat may produce a core radius which decreases 
with the halo circular velocity as $r_{\rm c} \propto V_{\rm h}^{-1/2}$.
This is not favored by our results. The predicted relation 
between $r_{\rm c}$ and $V_{\rm h}$ implies
that the halo concentration factor scales as
$r_{\rm h}/r_{\rm c}\propto V_{\rm h}^{3/2}$. Thus, at a given
redshift the concentration for a halo with 
$V_{\rm h}=250\kms$ is about 6 time larger than that
for a halo with $V_{\rm h}=75\kms$. From the results shown in 
Fig.\,\ref{fig_fig1}a and Fig.\,\ref{fig_fig2}a we see that the
resulting Tully-Fisher relation is much too shallow.

 If dark matter self-interaction (either scattering or annihilation)
is indeed responsible for the shallow profile of galactic haloes, then
the Tully-Fisher relation can be used to constrain 
the mass and cross section of dark matter particles.
Denote the mass and cross section by $m_X$ and
$\langle \sigma_{X}\vert v\vert\rangle$, respectively.
The collision rate per particle is 
$\Gamma=n_X\langle \sigma_{X}\vert v\vert\rangle$.
Collision is effective only in systems where 
$\Gamma^{-1}$ is smaller than the Hubble time.
This defines a critical density 
\beq
n_{\rm crit}={H(z)\over\langle\sigma_{X}\vert v\vert\rangle}\,,
\eeq     
above which the effect of self-interaction is important.
The above critical density defines a characteristic 
radius ($r_{\rm c}$) in a dark halo: $\rho (r_{\rm c})= 
m_X n_{\rm crit}$, where $\rho (r)$ is the halo density 
profile. Suppose that the halo profile before modification by 
self-interaction is
$\rho (r)=V_{\rm h}^2/(4\pi G r^2)$ near $r_{\rm c}$, the 
characteristic radius can be written  
\beq
r_{\rm c}={V_{\rm h}\over (4\pi G m_X n_{\rm crit})^{1/2}}
={V_{\rm h}\over [4\pi G H(z)]^{1/2}}
\left({\langle \sigma_X\vert v\vert\rangle\over m_X}\right)^{1/2}\,.
\eeq
If self-interaction of dark matter particles is to reduce 
the local density of dark matter particles, the characteristic
radius $r_{\rm c}$ may be identified as a `core' radius.
The halo concentration is then 
\beq
c\equiv {r_{\rm h}\over r_{\rm c}}
=\left[{\pi G\over 25 H(z)}\right]^{1/2}
\left({\langle \sigma_X\vert v\vert\rangle\over m_X}\right)^{-1/2}\,.
\eeq
Thus, if $\langle \sigma_X\vert v\vert\rangle$ is velocity
independent in the velocity range relevant for galactic haloes,
halo concentration $c$ is proportional to $1/\sqrt{H(z)}$, 
independent of $V_{\rm h}$. This is just the profile we want 
to explain the Tully-Fisher relation. In order to have a concentration
$c=7$ [assuming profile (\ref{profile})] at present time, the mass and 
cross section should satisfy 
\beq  
{\langle \sigma_X\vert v\vert\rangle\over m_X}
\sim 10^{-16} h^{-1} \left(c/7\right)^{-2}
{\rm \, cm^{3}\,s^{-1}\,GeV^{-1}}\,.
\eeq
The value of $\sigma_X$ implied is consistent with that 
obtained by Spergel \& Steinhardt (1999) and Firmani et al. (2000)
based on different arguments.
Much work remains to be done to see if a consistent model
can be found to fulfill the requirement. 

\section*{Acknowledgments}
We thank Gerhard B\"orner, Andi Burkert,
Ian Browne, Karsten Jedamzik, John McKean and Peter Wilkinson 
for helpful discussions and comments on the paper. We also thank our referee for 
a constructive report.

{}

\bsp
\label{lastpage}
\end{document}